\documentclass{article} % For LaTeX2e
\usepackage{iclr2027_conference,times}

\usepackage{amsmath,amsfonts,bm}

\def\eqref#1{equation~\ref{#1}}
\def\1{\bm{1}}

\DeclareMathAlphabet{\mathsfit}{\encodingdefault}{\sfdefault}{m}{sl}
\SetMathAlphabet{\mathsfit}{bold}{\encodingdefault}{\sfdefault}{bx}{n}

\usepackage{hyperref}
\usepackage{url}
\usepackage{graphicx}
\usepackage{amssymb}

\usepackage[most]{tcolorbox}
\usepackage{listings}

\definecolor{promptgray}{RGB}{247,247,247}
\definecolor{promptborder}{RGB}{190,190,190}

\newtcblisting{promptbox}[1][]{
  enhanced,
  breakable,
  colback=promptgray,
  colframe=promptborder,
  boxrule=0.6pt,
  arc=2pt,
  left=6pt,
  right=6pt,
  top=6pt,
  bottom=6pt,
  fonttitle=\bfseries,
  title=#1,
  listing only,
  listing options={
    basicstyle=\ttfamily\small,
    breaklines=true,
    breakatwhitespace=false,
    breakindent=0pt,
    breakautoindent=false,
    columns=fullflexible,
    keepspaces=false,
    showstringspaces=false
  }
}

\title{Vision-Language Agents for\\Active Perception in Optics Laboratories}

\author{
Ryan Lopez$^{1,\bigstar}$ \quad
Sachin Vaidya$^{1,2,3,\bigstar}$ \quad
Seou Choi$^{2}$ \quad
Serena Landers$^{1}$ \quad
Marin Soljačić$^{1,2,3}$
\\[0.5em]
$^{1}$Department of Physics, Massachusetts Institute of Technology\\
$^{2}$Research Laboratory of Electronics, Massachusetts Institute of Technology\\
$^{3}$NSF Institute for Artificial Intelligence and Fundamental Interactions\\[0.5em]
$^{\bigstar}$ denotes equal contribution.
\texttt{rnlopez@mit.edu}, \texttt{svaidya1@mit.edu}
}

\iclrfinalcopy % Uncomment for camera-ready version, but NOT for submission.
\begin{document}

\maketitle
\lhead{}

\begin{abstract}
Vision-language models (VLMs) are increasingly being used in scientific workflows, but their ability as agents to directly control laboratory experiments from visual feedback remains underexplored. This capability is important because many laboratory tasks do not naturally provide dense, pre-defined numerical objectives: informative signals can be sparse, intermittent, or visually ambiguous. A more general laboratory agent should instead be able to interpret visual observations, take actions to acquire useful feedback, and adapt its behavior based on the consequences of those actions. We study whether general-purpose VLMs can perform this kind of closed-loop scientific control using experimental optics as a testbed. We evaluate agents on three experimental systems that isolate distinct capabilities: a Michelson interferometer, a two-mirror cavity, and a four-mirror optical relay. The agents observe camera images, directly issue actuator and measurement commands, and retain their interaction history without receiving an engineered scalar objective during control. Across these experiments and matched simulations, we find that, given task-specific natural-language guidance, VLMs can estimate actuator-response relationships, resolve ambiguous observations through intervention, and actively create informative visual feedback when signals are sparse. These results suggest that pretrained multimodal models can serve as important decision-making agents within the experimental loop. Our work also establishes optics as a physically grounded testbed for visual reasoning and active perception in scientific agents. \href{https://anonymous.4open.science/r/vlm_active_perception-B115}{\textbf{\textcolor{blue}{Project Link.}}}
\end{abstract}

\section{Introduction}
Large language models are beginning to fundamentally change how experimental scientific work is carried out. They can generate analysis code, propose experimental steps, and interact with complex scientific software and instrumentation. However, many of these applications still place the model one step removed from the experiment itself. A model may decide what experiment should be performed or write code that implements a control routine, while the actual feedback loop is delegated to a task-specific program with a well-defined objective function for performing optimization. This division of labor is effective when the relevant state and objective can be specified in advance. Many experimental tasks, however, are not naturally presented in this form.

Real laboratory measurements are often incomplete and context dependent. A signal may be weak, obscured by another feature, or absent altogether; moreover, the meaning of an observation can depend on the sequence of actions that produced it. In such cases, useful feedback is not necessarily available at every step, and reducing an experiment to a ``fuzzy objective" requires encoding the domain knowledge and intuition that experts possess. Similar to human experimentalists, a more general scientific agent should be able to reason directly from experimental observations, use its own interventions to determine how the system responds, and, when necessary, take actions whose purpose is to obtain more informative measurements rather than immediately improve the final objective. Without this capability, scientific agents remain limited to executing experimental workflows whose relevant states, feedback signals, and objectives have already been specified by a human.

In this work, we ask whether general-purpose vision-language models (VLMs) can perform this kind of closed-loop scientific control. Our setting is experimental optics, which forms a backbone and an enabling technology across science and engineering. Optical alignment of a laser beam passing through multiple components is a familiar practical problem in this domain, and it provides a particularly useful setting for studying VLM-based agents. The underlying physics can be simple while the observations are not: laser beams can overlap, interfere, become difficult to distinguish from background features, or fail to reach a detector at all. Successful alignment therefore requires more than recognizing what is visible in a single image. An agent must infer how its controls affect the observed optical field, maintain information across a sequence of interventions, and sometimes create new observations before further progress is possible.

We construct three experimental tasks with varying demands on these capabilities to test pretrained VLMs: (1) a Michelson interferometer, (2) a two-mirror optical cavity, and (3) a four-mirror optical relay. In all three tasks, the agent receives only a natural-language description of the apparatus, goal, and general guidance and strategies, along with the camera images and interaction history generated by its previous actions. Based on these inputs, it returns actions that are autonomously executed in the laboratory using a robotic arm and motorized tools. We find that state-of-the-art VLMs show strong performance on these realistic sparse-feedback laboratory experiments. 

Our main contributions are summarized as follows. First, we demonstrate general-purpose VLMs operating directly in the closed loop of physical optical experiments using visual observations rather than engineered scalar objectives. Second, we introduce three experimental tasks that probe complementary capabilities: learning actuator-observation relationships through intervention, resolving ambiguous visual evidence using interaction history, and actively selecting measurements when useful feedback is absent. Third, we construct matched simulation environments that reproduce the principal laboratory behaviors and enable comparison with conventional numerical optimization baselines. Taken together, these results point towards a role for multimodal models that goes beyond proposing experiments or generating the code used to execute them. A scientific agent can instead help fully close the experimental loop, particularly in common sparse-feedback situations. The optical tasks explored here provide a concrete demonstration of this possibility and, more generally, a controlled physical setting for studying how agents reason and act when useful information must be discovered through interaction.

\begin{figure}[th]
\begin{center}
\includegraphics[width=\linewidth]{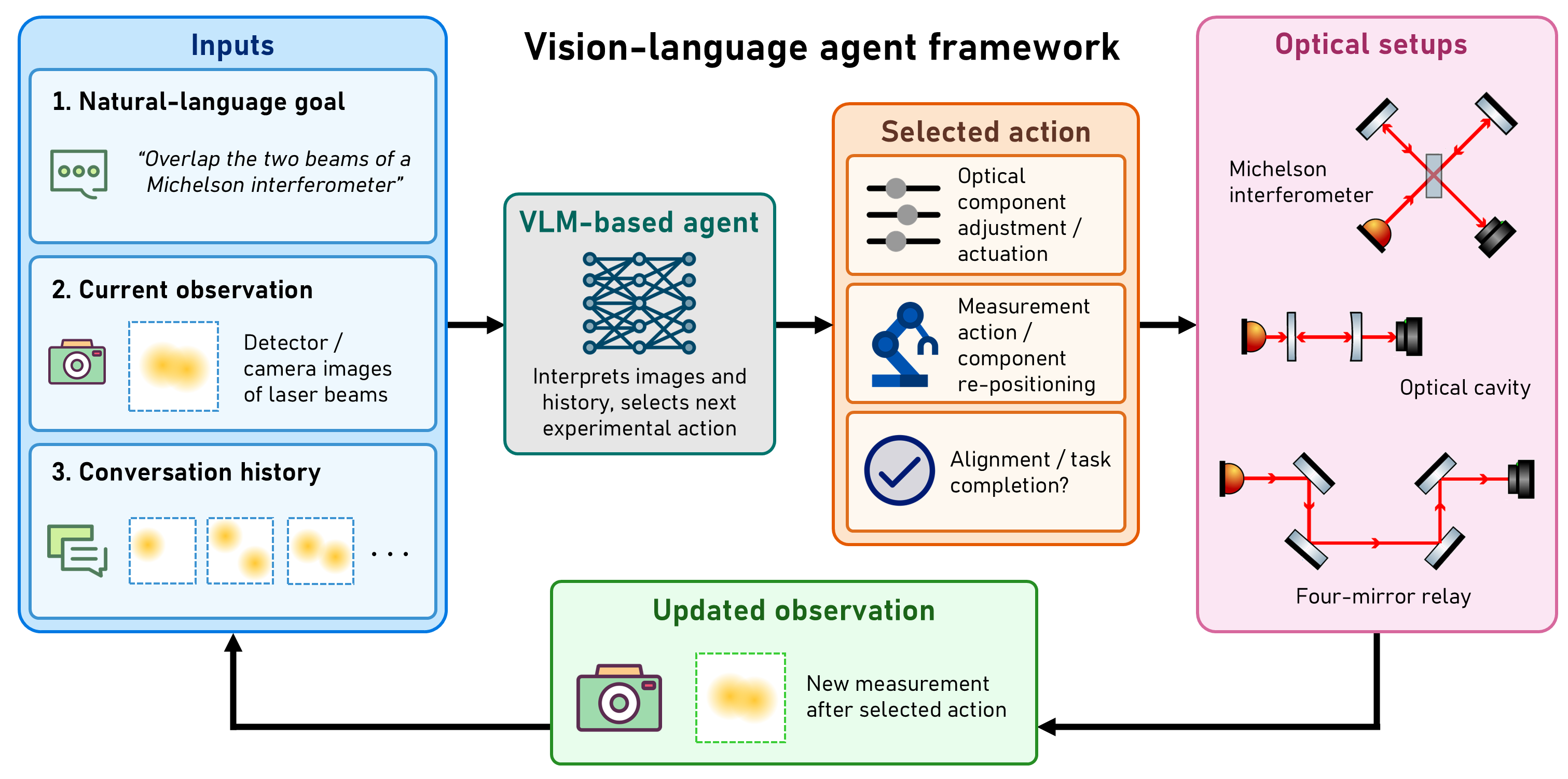}
\end{center}
\caption{\textbf{Vision-language agent framework for active perception.}
The agent receives an experimental goal in natural language, the current observation (such as a camera image), and the full conversation history which includes past observations. The model then selects an optical-control action, a measurement action, or a completion decision in a closed feedback loop with the experimental setup. The three chosen optical systems highlight complementary visual reasoning capabilities required for performing optical alignment.}
\label{fig:framework}
\vspace{-12pt}
\end{figure}

\section{Related Work}
Self-driving laboratories combine automated instrumentation with algorithmic experiment selection to iteratively pursue scientific objectives, and have been demonstrated across chemistry and materials discovery using robotic platforms and Bayesian or learned optimization strategies \cite{roch2018chemos,burger2020mobile,abolhasani2023rise}. More recently, large language models have been introduced as flexible reasoning and orchestration layers for scientific experimentation. Coscientist uses an LLM together with literature search, code execution, and laboratory APIs to design and execute chemistry experiments \cite{boiko2023autonomous}, while ChemCrow augments an LLM with domain-specific computational and experimental tools \cite{bran2024augmenting}. LLM agents have also been evaluated for autonomous atomic-force microscopy \cite{mandal2025afm}, and recent systems bring multimodal or agentic models directly into synchrotron experiments: Experiment Automation Agents use VLMs to operate materials-characterization workflows from multimodal inputs \cite{du2026experiment}, while an agentic X-ray scientist autonomously searches for and aligns diffraction reflections in both a virtual and physical beamline \cite{chen2026xray}. LLMs have also been applied directly to accelerator control, with~\cite{kaiser2025large} demonstrating that an LLM can tune a particle-accelerator subsystem from natural-language instructions. These works establish that foundation-model agents can plan and execute sophisticated physical experiments. We study a complementary question: whether a general-purpose VLM can use visual observations themselves as the principal feedback signal.

Optical systems have a long history of closed-loop alignment using engineered sensing and control signals. More recent approaches formulate beam tuning and alignment as numerical optimization problems. Bayesian optimization has been used for online image-based beam alignment in the SECAR recoil separator \cite{miskovich2022online} and, more generally, for autonomous tuning of X-ray and electron beamlines \cite{morris2024general}; the latter also evaluates the controller using a simulated digital twin. Reinforcement learning has likewise been applied to camera-guided laser alignment with motorized mirror mounts \cite{rakhmatulin2024reinforcement}, while AutoFocus combines real-time wavefront sensing, multi-objective Bayesian optimization, and digital-twin simulations for autonomous X-ray optical alignment \cite{rebuffi2025autofocus}. Derivative-free optimization methods have also been developed specifically for robot-controlled laboratory settings: CLUSTER accounts explicitly for the physical cost of changing experimental parameters and has been demonstrated on an optics experiment \cite{landers2026cluster}. These methods are effective when alignment quality can be reduced to a predefined error signal or numerical objective. We instead focus on regimes where constructing such an objective may be difficult or impossible.

Finally, robotic platforms for optical experimentation have also begun to address the physical manipulation layer required for more general laboratory autonomy. Recent work has introduced a robotics-for-optics platform that combines computer vision, precision robotic manipulation, and automated alignment for free-space optical experiments \cite{uddin2026robotics}. Building on this platform, subsequent work has demonstrated closed-loop robotic assembly, alignment, and recovery of precision optical systems, including autonomous construction of a tabletop laser cavity and recovery from induced misalignment \cite{choi2026framework}, as well as a programmable cloud-laboratory architecture that exposes optical experiments through a common software and control interface \cite{vaidya2026pico}. Together, these systems provide the actuation and experimental infrastructure on which more general perception-and-reasoning agents can operate.

\section{Methods and Results}

\subsection{Agent control protocol and experimental design}

We evaluate vision-language models as closed-loop controllers for three representative optical-alignment tasks: a Michelson interferometer, a two-mirror cavity, and a four-mirror relay. The framework is summarized in Fig.~\ref{fig:framework}. At the beginning of each alignment run, the model is initialized with a system prompt describing its role as an experimental agent, the optical apparatus and alignment objective, the available actions, and qualitative scientific guidance for interpreting observations, selecting alignment strategies, and verifying successful alignment. The agent then interacts with the experiment through a persistent conversation. At each turn, it produces structured outputs specifying actuator or measurement commands, working notes, and a decision on whether the task was complete. The requested action is autonomously executed in the laboratory, and the resulting camera image is returned to the model as the next observation. Because the full interaction history remains in context, the agent can jointly reason over previous observations and actions rather than treating each image independently. Complete prompts and output schemas are provided in the Appendix.

The agent is also provided with the permitted actuator ranges. Invalid or out-of-range commands are rejected and the model is prompted to issue a valid action. These actions are executed on the physical apparatus using Wi-Fi-connected motor controllers coupled to the mirror mounts and, for tasks requiring active signal acquisition, a robotic arm that can pick up and reposition a camera. The resulting control loop also naturally includes physical nonidealities such as motor slip, backlash, drift, and imperfect repeatability in camera placement, which the agent has to accommodate through subsequent visual observations.

For each experiment, we evaluate ten fixed initial misaligned optical configurations. Initial offsets are sampled uniformly from task-specific ranges and shared across models and simulation baselines, such that each method is evaluated from the same nominal set of starting conditions. These offsets are hidden from the agents and remain fixed throughout each run. The interaction budget is task dependent: 60 evaluations for the Michelson interferometer, 50 for the two-mirror cavity, and 100 for the four-mirror relay. Each evaluation consists of one agent action followed by the resulting observation of a camera beam image. A run terminates when the agent declares the alignment complete or when the evaluation budget is exhausted. Importantly, an agent's completion declaration is not itself used to determine success; terminal alignment quality is assessed independently using the specific metrics defined below.

Across the three experiments, we primarily evaluate GPT-5.6 Sol and Gemini 3.5 Flash. For selected tasks, we additionally test other models, including Claude Sonnet 5 and Qwen3-VL 32B Instruct, to provide further points of comparison. All models are accessed through OpenRouter. Model identifiers and inference settings are provided in the Appendix.

\subsection{Inferring actuator-response relationships:  Michelson interferometer}

\begin{figure}[t]
\begin{center}
\includegraphics[width=\linewidth]{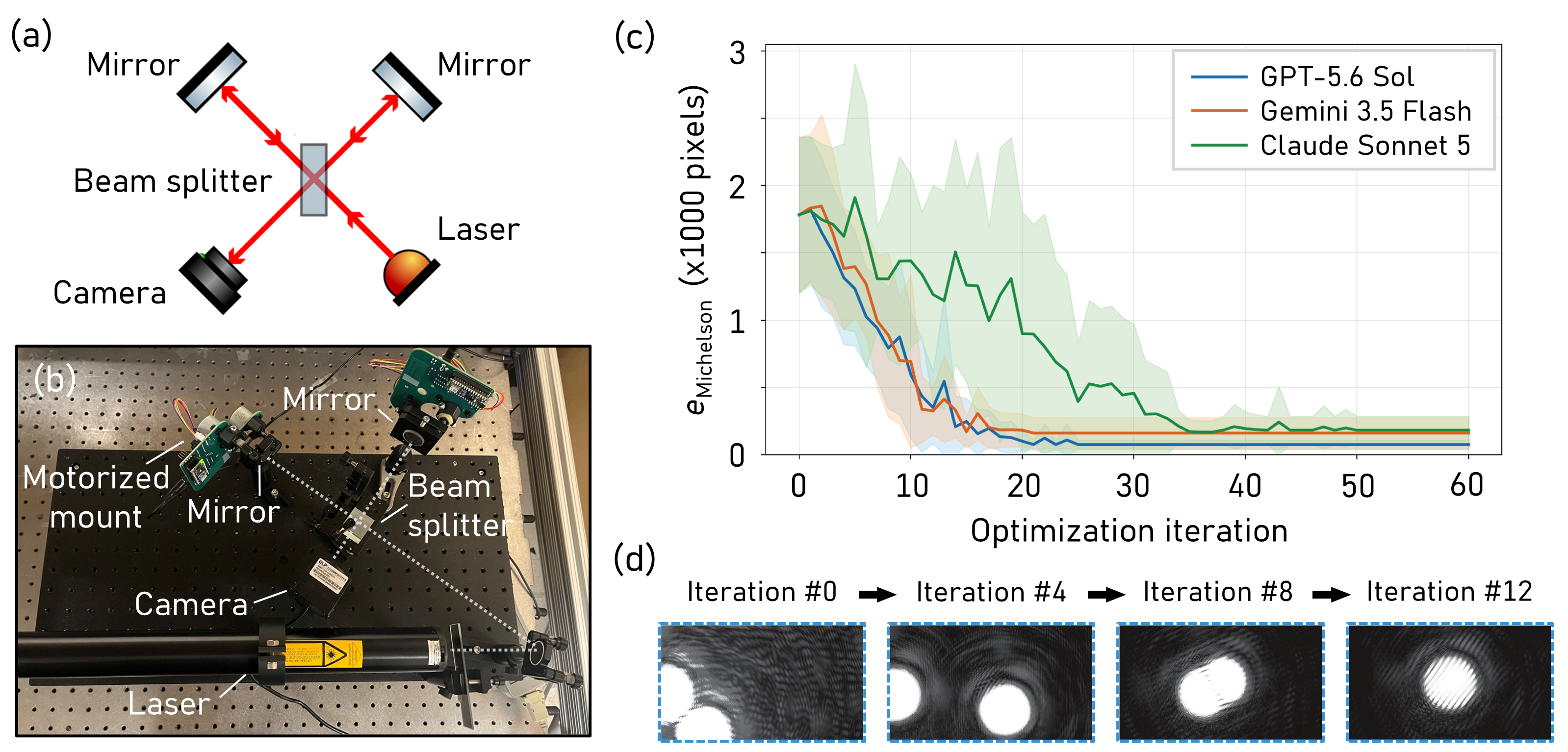}
\end{center}
\caption{\textbf{Vision-language-agent alignment of a Michelson interferometer.} (a) Schematic of the Michelson interferometer setup. (b) Photograph of the corresponding laboratory setup with motorized mirror mounts used for closed-loop alignment. The white dotted line shows the intended laser beam path. (c) Mean summed distance of each beam from the center of camera over successive optimization iterations for three VLMs; shaded regions indicate the sample standard deviation across repeated runs. (d) Camera images during a representative GPT-5.6 Sol run, progressing from separated, clipped beam patterns to a centered overlap with linear interference fringes.}
\label{fig:michelson_lab}
\vspace{-12pt}
\end{figure}

In this first experiment, we test whether a VLM-based agent can visually interpret an optical state and learn how its available actuators affect that state through interaction. The Michelson interferometer setup consists of a HeNe laser at 633nm, a beamsplitter, two motorized mirrors, and a beam detection camera. The beamsplitter divides the incident beam into two arms, which reflect from their respective mirrors and recombine before reaching the camera (Fig.~\ref{fig:michelson_lab}a,b). Each mirror provides two angular control degrees of freedom, controlled via motorized mirror mounts. The alignment objective is to center both beams on the camera and bring them into overlap.

This task tests visual interpretation together with identification of actuator effects. Depending on the initial misalignment, the two beams can appear as well-separated spots, partially overlapped, clipped by the camera boundary, or lie entirely outside the camera's field of view. Near alignment, interference further obscures the individual beam profiles, making their centers difficult to infer directly from a single image. Although the prompt describes the optical layout and available controls, it does not provide the mapping from motor commands to image-space beam motion. The agent therefore has to infer these relationships through interaction. It is instructed to associate visible features with actuators using controlled perturbations and to verify candidate alignments using observations on both sides of the inferred target (i.e., slightly adjust the actuators in both directions to ensure alignment was actually reached) with perturbation-and-return checks.

To characterize the apparatus independently for initialization and scoring, we calibrate each mirror's image-space response. One beam is held near the camera center while the other mirror is scanned over a grid of motor positions. Beam centers are extracted from the resulting images using a vision-language model and manually verified, and a linear model is fit between motor coordinates and image-space beam position. The procedure is then repeated for the second mirror. Initial misalignments are generated by independently sampling horizontal and vertical motor step offsets from $\pm 6144$ and $\pm 4096$ respectively, where one full motor revolution is $4096$ steps. These ranges are chosen such that one or both beams can potentially lie outside the camera field of view at the start.

We define the alignment error as the summed Euclidean distance of the two beam centers from the camera center,
$
e_{\mathrm{Michelson}}
=
\left\|\mathbf r_1-\mathbf r_c\right\|_2
+
\left\|\mathbf r_2-\mathbf r_c\right\|_2,
$ where $\mathbf r_1$ and $\mathbf r_2$ denote the two beam-center positions and $\mathbf r_c$ denotes the center of the camera's field of view, all expressed in pixels. For evaluation, beam positions are inferred from the commanded motor coordinates using the independently measured calibration. This provides a scoring metric even when a beam lies outside the camera field of view or when interference makes its center difficult to identify directly from the recorded image. This metric is used only for initialization and evaluation and is not provided to the agent.

\textbf{Results --} For this task, we tested GPT-5.6 Sol, Gemini 3.5 Flash and Claude Sonnet 5, with the same prompt. All three models tested successfully aligned the Michelson interferometer from the initial misalignments. All three models declared completion in all 10 laboratory runs, and manual inspection of the terminal images confirmed that the beams were brought into overlap near the camera center. The precision of the final alignment, however, differed substantially across models (Fig.~\ref{fig:michelson_lab}c). GPT-5.6 Sol consistently produced near-complete overlap and centering (e.g., Step 12 in Fig.~\ref{fig:michelson_lab}). Gemini 3.5 Flash and Claude Sonnet 5 also consistently brought the beams towards the camera center, but often terminated with imperfect overlap. In the least precise cases, the combined beam remained visibly elongated, with an appearance comparable to Step 8 in Fig.~\ref{fig:michelson_lab}d.

The metric defined above provided a quantitative measure of terminal alignment quality. The mean terminal alignment error was 75 pixels for GPT-5.6 Sol, 161 pixels for Gemini 3.5 Flash, and 182 pixels for Claude Sonnet 5, for images of size $1920\times1080$ pixels. The models also differed in the number of observations used before declaring completion. GPT-5.6 Sol terminated after a median of 19 evaluations, compared with 16 for Gemini 3.5 Flash and 27.5 for Claude Sonnet 5.

A GPT-5.6 Sol trajectory shown in Fig.~\ref{fig:michelson_lab}d illustrates a typical run. Starting from two separated spots, the agent varied one axis of the first mirror while holding the remaining controls fixed and identified which visible beam responded. From successive observations, it inferred the sign and approximate sensitivity of the horizontal and vertical actuator responses. It then used observations on opposite sides of the camera center to interpolate towards a centered motor setting. With the first beam fixed, it repeated the procedure for the second mirror and brought the second beam onto the centered reference. Near overlap, the individual beam centers became visually ambiguous due to interference. The agent did not terminate solely because the current image resembled an aligned interference pattern. Instead, it deliberately displaced each beam until it became separately visible and then returned the corresponding actuator to the inferred aligned setting, verifying that the overlap was recovered. This sequence combined intervention, observation history, and explicit verification: the model first identified the approximately linear actuator-observation relationships and then used those inferred relationships to test its own candidate solution.

\subsection{Resolving ambiguous visual feedback: Two-mirror cavity}

\begin{figure}[t]
\begin{center}
\includegraphics[width=\linewidth]{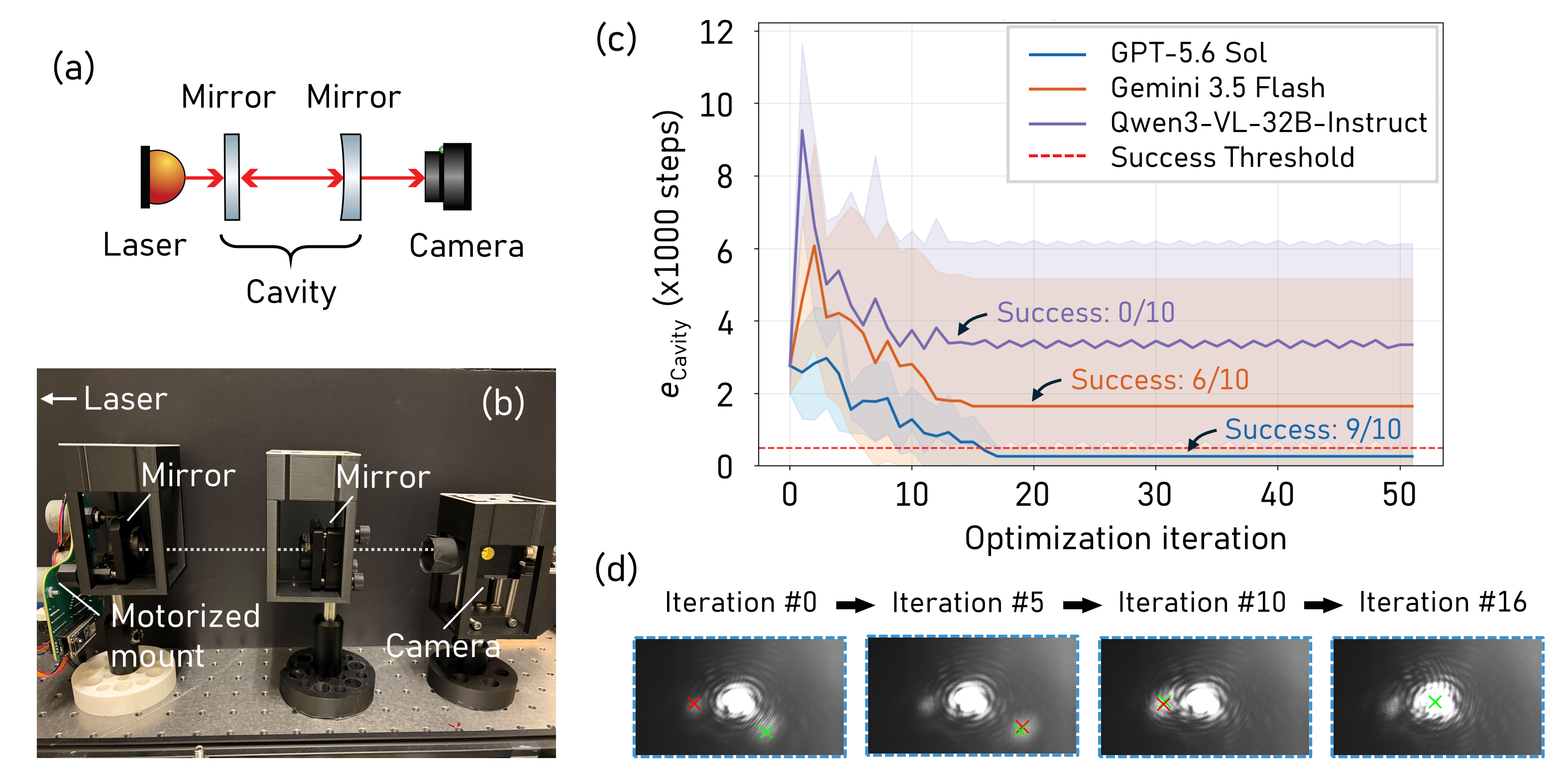}
\end{center}
\caption{\textbf{Vision-language-agent alignment of a two-mirror cavity.} 
(a) Schematic of the two-mirror cavity, showing the laser, in-coupling mirror, out-coupling mirror, and a camera. (b) Photograph of the corresponding laboratory setup with a motorized mirror mount attached to the in-coupling mirror used for closed-loop alignment. The white dotted line shows the intended laser beam path. (c) Mean distance measured in motor steps between the center of the secondary beam and the main beam over optimization iterations for three VLMs; shaded regions indicate the sample standard deviation across repeated runs. (d) Camera images during a representative run. The green $\times$ marks the true location of the secondary beam and the red $\times$ marks where the agent thinks the beam is. These marks are for post-hoc analysis only; they were not provided to the agent.}
\label{fig:cavity_lab}
\vspace{-12 pt}
\end{figure}

In this second experiment, we test whether an agent can resolve visually ambiguous feedback by reasoning over the history of its actions and observations. The two-mirror cavity setup consists of a laser diode (808nm wavelength), two mirrors (an in-coupler and out-coupler), and a beam detection camera positioned to observe the transmitted light, as shown schematically in Fig.~\ref{fig:cavity_lab}a. The camera records a bright primary beam together with a weaker secondary beam that completes an additional round trip between the cavity mirrors. The out-coupler mirror remains fixed, while two motorized angular controls on the in-coupler mirror translate the secondary beam across the camera image while leaving the primary beam approximately stationary. The alignment objective is to bring the secondary beam into overlap with this fixed primary beam, which ensures multiple passes of the laser beam within the cavity.

This task requires distinguishing a weak signal of the secondary beam in the presence of background noise and the much stronger signal of the primary beam in the same visual field. Furthermore, this task introduces an ambiguity in the interpretation of a disappearing signal. The secondary beam becomes difficult to distinguish when it overlaps the much brighter primary beam, but it can also disappear by moving outside the camera's field of view. Its absence from a single image does not by itself establish successful alignment. Resolving this ambiguity requires reasoning over the relationship between previous actuator commands and observations. The prompt explicitly acknowledges this ambiguity and instructs the agent to use command-correlated motion, observation history, and controlled probes around a candidate overlap position to verify alignment.

The prompt specifies the overlap objective, absolute motor limits, and the correspondence between motor axes and image axes, but requires the agent to infer the response sign and sensitivity of each actuator from observed beam motion. It further prescribes an approximately linear, independent-axis calibration strategy, rejection of stationary optical artifacts, and two-sided verification around a candidate overlap position. When close to alignment, the agent is instructed to establish that nearby perturbations cause the beam to reappear on opposite sides of the primary beam. If the evaluation budget is reached before this verification is complete, the agent is instructed to return the best-supported motor configuration.

We define the alignment error as the Euclidean distance from the aligned reference in motor space,
$
e_{\mathrm{cavity}}
=
\sqrt{
(u_x-u_x^\star)^2 +
(u_y-u_y^\star)^2
},
$ where $(u_x,u_y)$ denote the current motor positions and $(u_x^\star,u_y^\star)$ denote the aligned reference positions. A terminal configuration is considered successful when $e_{\mathrm{cavity}} \leq 500$ motor steps, which corresponds approximately to the center of the secondary beam lying within the spatial profile of the primary beam. As before, this metric is used for evaluation of model performance and is not provided to the agent.

\textbf{Results -- } Under the success criterion defined above, GPT-5.6 Sol successfully aligned 9/10 laboratory runs (Fig.~\ref{fig:cavity_lab}c), while Gemini 3.5 Flash succeeded in 6/10. We also tested a smaller pretrained VLM, Qwen3-VL-32B-Instruct, which failed all 10 attempts. All three models nevertheless declared the task complete and terminated early in every run, indicating that self-reported completion alone was not a reliable measure of alignment. The mean terminal errors were 266, 1,649, and 3,345 motor steps for GPT-5.6 Sol, Gemini 3.5 Flash, and Qwen3-VL, respectively.

A representative successful GPT-5.6 Sol trajectory, shown in Fig.~\ref{fig:cavity_lab}d, illustrates how the model resolves this ambiguity using its interaction history. Early in the run, the agent incorrectly identified a stationary background feature as the secondary beam. Subsequent motor adjustments showed that this feature did not move with the controlled mirror, allowing the agent to reject it and identify the actual secondary beam. Once the beam was spatially separated from the bright primary beam and could be localized more reliably, GPT-5.6 Sol used several observations to estimate an approximately linear relationship between motor commands and image-space beam position. It refined this relationship with additional probes and interpolated the motor settings to bring the secondary beam into overlap with the primary beam. The agent also probed motor positions on either side of the candidate along both axes and verified that the secondary beam reappeared on opposite sides of the primary beam. Only after establishing this did it return to the predicted overlap configuration and declare completion.

The oscillatory trend observed for Qwen3-VL-32B-Instruct in Fig.~\ref{fig:cavity_lab}c was due to a single run in which the model failed to identify the secondary beam and cycled through the same commands repeatedly. The least successful Gemini 3.5 Flash runs were dominated by errors in visual interpretation. In these cases, incorrect identification of the weak secondary beam led to inaccurate actuator-response estimates and, consequently, incorrect completion decisions. Thus, although the task involved only two control degrees of freedom, reliable performance depended strongly on maintaining the identity and motion of a weak visual feature across multiple interventions.

\subsection{Active measurement selection under missing feedback: Four-mirror relay}

\begin{figure}[t]
\begin{center}
\includegraphics[width=\linewidth]{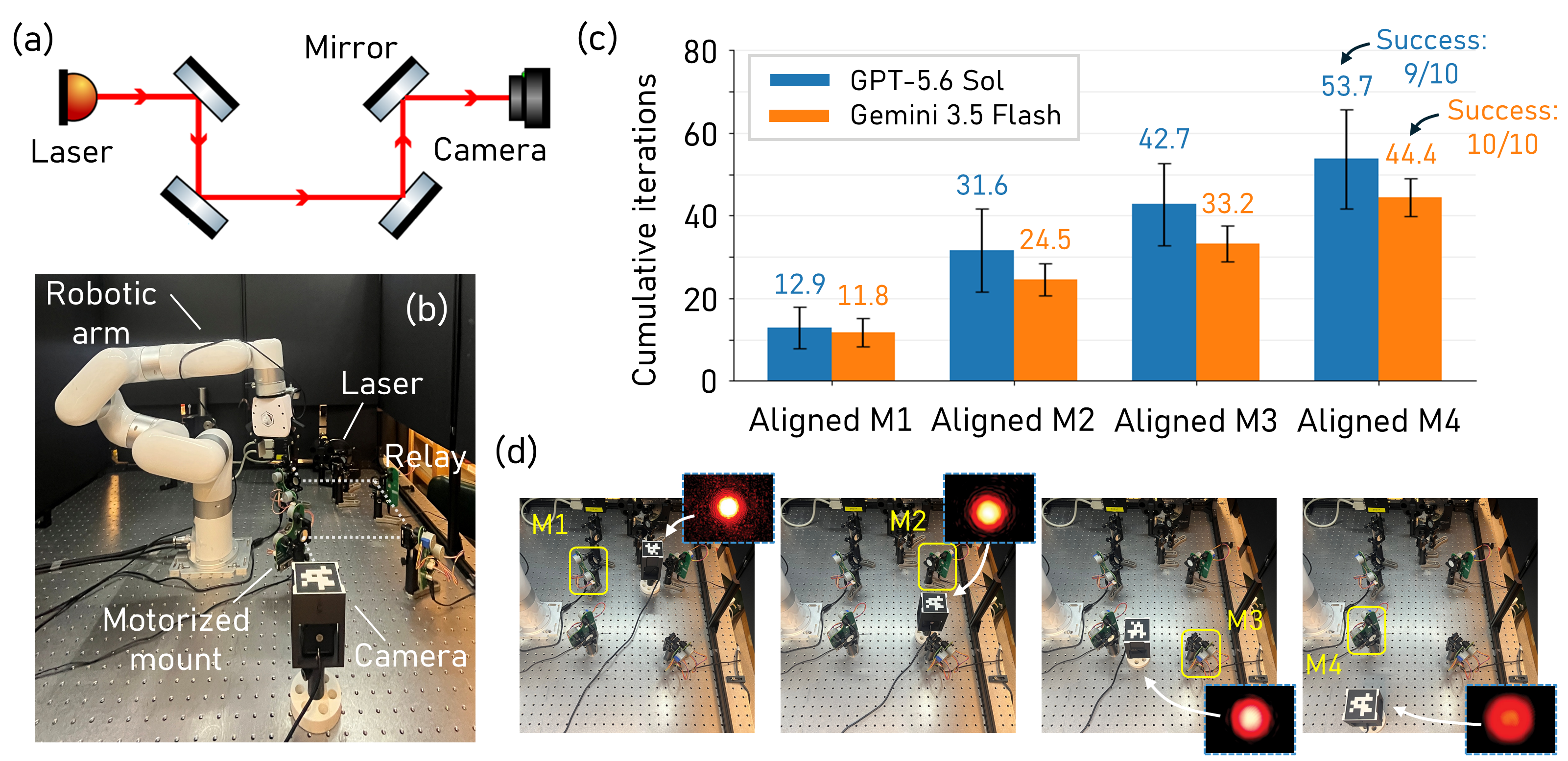}
\end{center}
\caption{\textbf{Vision-language-agent alignment of a four-mirror relay.} (a) Schematic
of the four-mirror relay, showing the laser, four mirrors, and a camera. (b) Photograph of the corresponding laboratory setup with a robotic arm for pick-and-place and motorized mirror mounts attached to each mirror. The white dotted line shows the intended laser beam path. (c) The mean cumulative number of iterations performed to align each mirror across the two VLMs tested, measured from the start of each run. (d) Top-down views of the four-mirror relay showing typical camera placements (and beam images) at each stage, with the mirror being aligned highlighted in yellow.}
\label{fig:relay_lab}
\vspace{-12pt}
\end{figure}

In this final experiment, we test whether an agent can actively acquire informative observations when the available visual feedback is insufficient to determine a corrective action. The four-mirror relay setup forms an approximately $8$-inch by $11$-inch rectangular beam path on the laboratory table, as shown schematically in Fig.~\ref{fig:relay_lab}a. Each mirror has two motorized angular controls, giving eight optical degrees of freedom in total. The intended beam path reflects successively from all four mirrors (labeled M1 to M4) before reaching a final observation location.

Initial misalignments of the mirrors can cause the beam to miss one or more downstream mirrors entirely, leaving a camera placed at the end of the relay with no visible signal. Alignment of this setup requires the agent to recover feedback at intermediate locations along the optical path. A natural strategy is to observe and align successive propagation segments while moving the observation point downstream. The task places several coupled demands on the agent: it must interpret visible, clipped, and absent beam signals, infer how mirror adjustments affect beam propagation, and choose camera locations that expose otherwise unobserved alignment errors. When a downstream observation is blank, the agent must first determine where to acquire useful visual feedback before it can decide on an appropriate corrective action.

At each iteration, the agent can either adjust the two angular controls of a single mirror or move the camera to one of eight predefined observation locations. Eight camera locations are available, two along each of the four propagation segments: one closer to the upstream mirror and one farther downstream, towards the receiving mirror or final output location. After either action, the agent receives a single camera image from its current observation location. When a camera reposition action is chosen by the agent, a robotic arm picks up and places the camera at the requested location using the methods developed in~\cite{uddin2026robotics} (Fig.~\ref{fig:relay_lab}b). Mirror adjustments use the Wi-Fi-connected motor controllers described above.

The prompt specifies the overall goal of propagating the beam through the relay and centering it in the final camera image, but does not prescribe an alignment order or a sequence of camera placements. The agent determines both where to observe the system and when to adjust each mirror. The prompt also describes the relay geometry, the eight available camera placement locations, motor limits, and the nominal actuation scale. It instructs the model to estimate actuator-response from observations, account for imperfect motor actuation and camera placement, and distinguish localized beam spots from clipping, speckle, and missing signals. Completion of this task requires a centered, localized beam at the final measurement location after passing through all four mirrors.

\textbf{Results -- } To quantify alignment progress, we defined milestones: a mirror was considered aligned once the beam was centered at the farther camera position along its outgoing segment and the agent made no subsequent adjustments to that mirror. For this experiment, we tested two models: GPT-5.6 Sol and Gemini 3.5 Flash. GPT-5.6 Sol successfully aligned 9/10 hidden initial configurations, although it declared success in all ten runs. Gemini 3.5 Flash successfully aligned 10/10 configurations. Figure~\ref{fig:relay_lab}c shows the average number of iterations required to reach the alignment milestones.

Each run began with an observation at the final camera position, which showed no signal for all tested initial conditions. The agents generally responded by moving the camera upstream to the beginning of the relay, where useful optical feedback could be recovered. From there, they progressively moved the observation point downstream, adjusting the corresponding mirror until the beam was centered before advancing to the next segment. Thus, rather than attempting to optimize directly from the uninformative final observation, the agents decomposed the relay into a sequence of locally observable alignment problems (Fig.~\ref{fig:relay_lab}d).

The agents also used camera backtracking when downstream feedback was lost. Among successful runs, GPT-5.6 Sol moved the camera back to an earlier observation point in four trajectories, while Gemini 3.5 Flash did so in two. These events typically occurred when the signal disappeared at the farther camera position, prompting the model to return to the nearer observation point to refine alignment before proceeding. GPT-5.6 Sol's single failed run arose from an error in visual interpretation at the first observation position. The agent failed to localize a distinct beam and instead observed only diffuse red speckle, which it treated as the signal. This error propagated through the trajectory, and the model ultimately declared success while observing only a diffuse red image.

\subsection{Simulation and numerical baselines}

We construct matched optical simulations of all three optical systems to reproduce the observation and control structure of the corresponding laboratory experiments. Vision-language agents interact with these environments through the same action interfaces used in the laboratory. We additionally compare against Bayesian optimization (BO) and random search. BO receives a privileged scalar objective computed from the simulator state, whereas the vision-language agents receive only rendered camera images, interaction history, and the natural-language task description. Full simulation parameters, optimizer settings, and evaluation details are provided in the Appendix.

The simulations reproduced the principal laboratory convergence trends for the agents. The numerical baselines exhibited task-dependent behavior: BO was effective when supplied with an informative scalar objective and, in the case of the two-mirror cavity, ultimately refined the alignment more precisely than the vision-language agent. Its performance degraded when the objective provided little information about intermediate progress. This was most apparent in the four-mirror relay, where BO failed to progress beyond the third mirror because configurations in which the beam did not reach the final detector received the same objective value. The agent instead acquired intermediate observations by moving the camera and decomposed the task into sequential alignment steps. Detailed simulation-to-laboratory comparisons and baseline results are reported in the Appendix.

More broadly, the agreement between simulation and experiment that we observe suggests that these environments can serve as practical testbeds for developing laboratory agents before physical deployment. The relatively small simulation-to-laboratory gap observed here makes it possible to evaluate prompting strategies, refine agent behavior, and potentially fine-tune models through large numbers of inexpensive simulated interactions before transferring them to hardware. Simulation also makes it straightforward to vary hidden system states, perturbations, noise levels, and task difficulty beyond what is practical to evaluate repeatedly in the laboratory. This points to a second use of these environments: sufficiently diverse and challenging optical simulations could provide a physically grounded benchmark for visual reasoning and active perception in scientific tasks. Developing such training environments and benchmarks at scale is left for future work.

\section{Conclusions}
We showed that pretrained VLMs can act directly within an experimental feedback loop, using visual observations not only to optimize a system but also to determine which observations are informative enough to support control. Across the three optical systems, the VLM-based agents estimated actuator-observation relationships from their own interventions, used observation history to resolve ambiguous visual evidence, and, when downstream feedback disappeared entirely, actively changed where they measured the system to recover informative observations. These capabilities were sufficient to close the loop in laboratory experiments despite the presence of background noise, interference, weak signals, mechanical nonidealities, and missing feedback.

Our work suggests a broader role for multimodal foundation models in scientific automation. Rather than replacing conventional controllers and optimizers, VLM agents may be most useful at the layer where perception, measurement selection, and control strategy cannot easily be reduced to a numerical objective, such as in sparse-signal scenarios. Once informative feedback has been established, standard optimization methods can remain valuable for refinement. The agreement we observe between optical simulations and laboratory behavior further makes this setting useful, where simulations can provide scalable environments for evaluating and potentially training agents on active perception in scientific tasks.

\section*{Acknowledgment}
S.V. thanks Logan G. Wright for discussions on this topic. R.L. acknowledges support from the Dean of Science Fellowship. S.C. acknowledges support from the Korea Foundation for Advanced Studies Overseas PhD Scholarship. S.V. and M.S. acknowledge support from NSF under Cooperative Agreement PHY-2019786 (The NSF AI Institute for Artificial Intelligence and Fundamental Interactions). This work was also supported in part by the U.S. Army DEVCOM ARL Army Research Office through the MIT Institute for Soldier Nanotechnologies under Cooperative Agreement W911NF-23-2-0121, the MIT Generative AI Impact Consortium (MGAIC), and Shell International Exploration and Production Inc.

\section*{AI use statement}

In this work, we used generative AI tools to create and edit software code, analyze existing literature, and copyedit the manuscript to improve readability. Generative AI was also used to refine the language in the final prompts provided to the VLM-based agents. The tools and models used were GPT-5.6 Sol and OpenAI Codex. We have not used generative AI tools to help develop theoretical models or conceptual frameworks, propose or refine hypotheses, design or provide feedback on research  methodology or experiments, support qualitative and thematic data analysis, or interpret results. We have reviewed all AI-assisted work: software code was read through and tested, literature search results were checked, and only minor readability editing was done. The authors take full responsibility for the final content of this work, including text, claims or artifacts produced with the aid of generative AI.

\bibliography{iclr2027_conference}
\bibliographystyle{iclr2027_conference}

\clearpage

\appendix
\begin{center}
    {\LARGE \textbf{Appendices}}
\end{center}
\vspace{1em}

\section{Simulation environments and numerical baselines}

We construct matched simulations for each optical-alignment task that reproduce the observation and control structure of the corresponding laboratory experiment. Rather than attempting to model the full optical system at high physical fidelity, the simulations capture the visual features, actuator responses, field-of-view limitations, and signal-loss mechanisms that are relevant to the agent's decision making. The vision-language agents interact with the simulated systems through the same action interfaces used in the laboratory experiments and receive one rendered camera image after each evaluation.

For the Michelson interferometer, we model the two returning beams as Gaussian optical fields whose widths and motor-dependent positions are matched to the laboratory apparatus. The simulated camera image is generated from their coherent superposition, producing interference in regions where the two fields overlap. A fixed relative spatial phase gradient reproduces the approximate orientation and spacing of the interference fringes observed experimentally. This lightweight model captures the principal visual regimes encountered during alignment, including separated beams, clipping at the image boundary, partial overlap, and centered interference, without requiring a full electromagnetic simulation.

For the two-mirror cavity, we construct the observation model from representative laboratory beam profiles. The primary- and secondary-beam intensity distributions are extracted from laboratory images, with the primary profile held fixed and the secondary profile translated according to the experimentally calibrated motor response. The rendered camera image is formed by summing the two intensity profiles. This approximation preserves the characteristic beam shapes, relative brightness, and visual ambiguity that occurs when the weak secondary beam approaches the bright primary beam, while omitting phase-dependent interference.

For the four-mirror relay, we use three-dimensional geometric ray tracing with the same nominal optical layout, circular mirror apertures, camera field of view, and eight observation locations as the laboratory apparatus. Mirror commands modify the reflected ray directions, and the rendered camera image depends on whether and where the propagated beam intersects the selected observation plane. The simulation therefore reproduces the loss of downstream signal when the beam misses an intermediate mirror or camera. As in the laboratory, the agent can either adjust a mirror or reposition the camera at each evaluation and receives an image only from the currently selected observation location.

To provide numerical optimization baselines, we additionally evaluate Bayesian optimization (BO) and random search in simulation. BO uses a Gaussian-process surrogate with a Matérn-5/2 kernel and selects successive actuator configurations using an expected-improvement acquisition function with $\xi=0.01$. For the Michelson interferometer and two-mirror cavity, the objective is the negative of the corresponding alignment error defined in the main text. For the four-mirror relay, BO maximizes the negative pixel distance between the beam and center of the final fixed camera. Configurations for which the beam does not reach the final camera are assigned the same objective value of $-10^6$. Random search independently samples actuator configurations uniformly over the permitted ranges. The numerical baselines use the same hidden initial offsets, actuator bounds, and evaluation budgets as the vision-language agents. 

The available feedback differs by design. BO receives a privileged scalar objective computed from the simulator state, which in a physical experiment could require substantial task-specific processing to construct robustly. The vision-language agents instead receive only rendered camera images, their interaction history, and a natural-language description of the task and apparatus. The comparison contrasts numerical search supplied with a predefined measure of progress against agents that must determine useful feedback from visual observations and, when necessary, actively acquire new observations.

\begin{figure}[ht]
\begin{center}
\includegraphics[width=\linewidth]{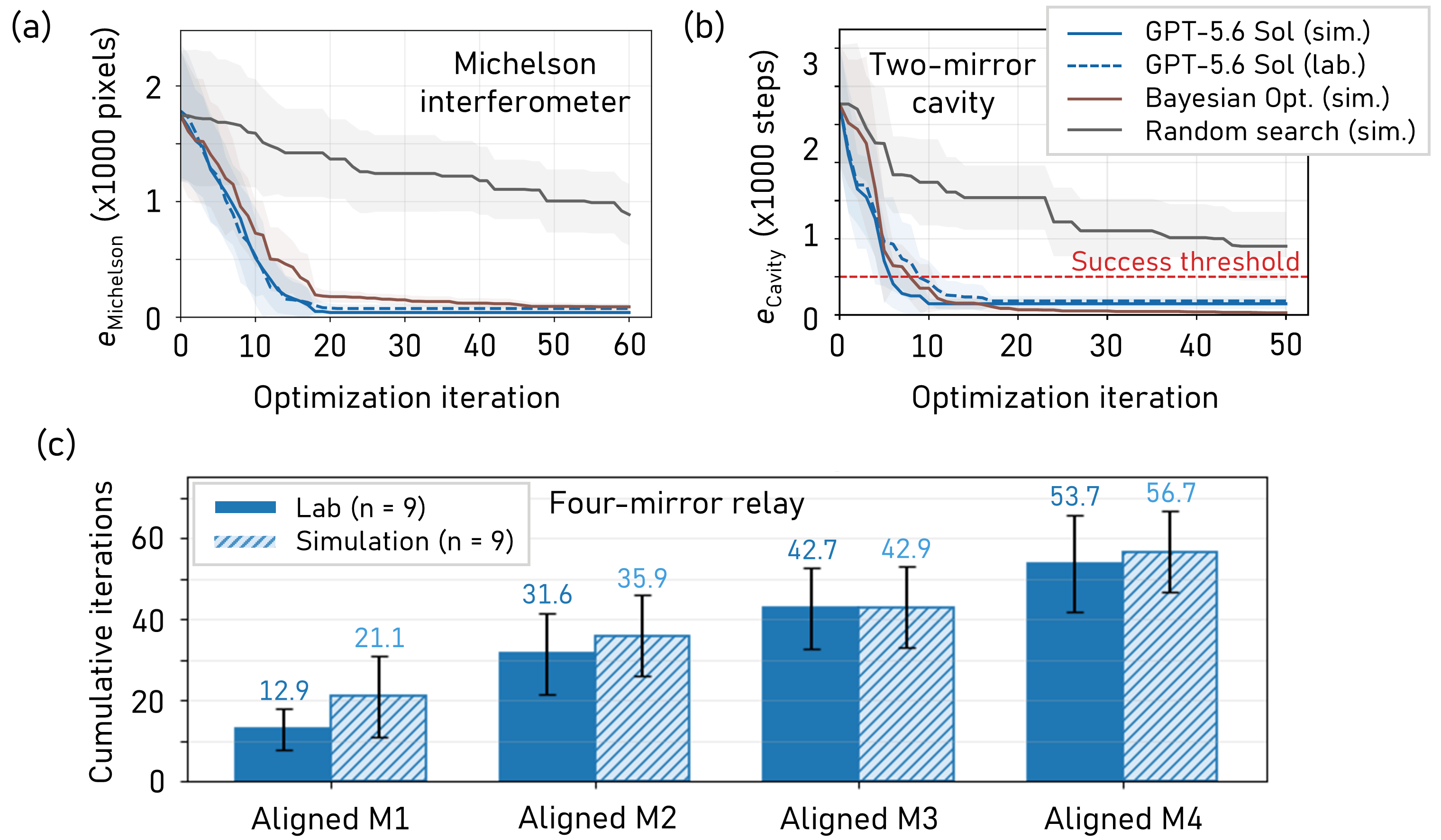}
\end{center}
\caption{\textbf{Simulation reproduces laboratory alignment behavior and enables comparison with privileged optimization baselines.} (a) Mean best-so-far Michelson interferometer alignment error, using the same error metric defined in the main text. Simulated and laboratory GPT-5.6 Sol trajectories show similar convergence, while GPT-5.6 Sol reaches a lower final error than BO and random search. (b) Mean best-so-far two-mirror cavity alignment error, using the same error metric and success threshold defined in the main text. Simulated and laboratory GPT-5.6 Sol trajectories again show similar convergence, while BO converges more slowly initially but ultimately reaches lower error. In (a-b), best-so-far error is shown because it is the natural performance measure for BO and random search. Shaded regions indicate sample standard deviation across runs. (c) Mean cumulative number of iterations, measured from the start of each run, required for GPT-5.6 Sol to align each successive mirror in the four-mirror relay in laboratory and simulation experiments; error bars indicate the sample standard deviation. The simulated and laboratory trajectories show similar sequential alignment behavior. Numerical baselines are not shown in this panel because all BO and random search runs failed to complete the relay: neither method propagated the beam to the fourth mirror, and BO therefore never obtained an informative final-camera signal from which to optimize. The laboratory results include the nine successful runs, and the simulation comparison uses the corresponding nine initial configurations.
}
\label{fig:combined_results}
\end{figure}

\textbf{Results --} We evaluate GPT-5.6 Sol in simulation using the same prompts and control interfaces as in the laboratory experiments. All simulated methods, including GPT-5.6 Sol, BO, and random search, are initialized from the same nominal hidden motor offsets used in the corresponding ten laboratory runs. For the four-mirror relay, only the nine initial configurations corresponding to the successful laboratory runs were used. Across all three optical systems, the simulated and laboratory experiments converged on similar scales and showed similar alignment behavior (Fig.~\ref{fig:combined_results}).

For the Michelson interferometer and two-mirror cavity, GPT-5.6 Sol and BO both reached low-error alignments on similar iteration scales, while random search performed substantially worse. This is notable because BO receives a privileged scalar objective computed from the simulator state, whereas GPT-5.6 Sol acts from raw visual feedback. However, the agent can exploit additional structure provided through the prompt and visual observations, including which controls affect particular optical features and the approximately linear response of the system. Through controlled perturbations, it can estimate actuator-response directions and sensitivities and then interpolate towards an aligned configuration. Consequently, its behavior can be viewed as structured, vision-based system identification rather than generic black-box optimization.

The contrast was much stronger in the four-mirror relay. With the camera fixed at the final observation location, BO receives no information from the numerical objective until the beam has propagated through the relay and reached the final camera. Under the tested search budget, no BO or random-search run propagated the beam beyond the third mirror. BO remained on the same flat objective value for the entire search. GPT-5.6 Sol instead repositioned the camera to recover intermediate visual feedback and progressively aligned successive sections of the relay. In this setting, actively acquiring informative observations was essential: searching actuator configurations from the final camera alone was unlikely to encounter a state that provided useful feedback.

Taken together, these results suggest that the value of the vision-language agent is not that it is uniformly a better numerical optimizer. When an informative scalar objective is available, BO can achieve comparable performance. In practice, however, such privileged objectives may be unavailable or difficult to engineer from raw experimental measurements. Moreover, even a well-defined final objective may provide little useful feedback when the system is far from the desired state. In these settings, a vision-language agent can use contextual information from the experiment and actively acquire informative signals needed to make progress.

\section{Model Identifiers and Inference Settings}

All models were accessed through OpenRouter using default inference settings unless otherwise specified. Laboratory experiments used GPT-5.6 Sol (openai/gpt-5.6-sol; high reasoning effort; 20,000 maximum output tokens for the Michelson interferometer and two-mirror cavity, and 12,000 tokens for the four-mirror relay), Gemini 3.5 Flash (google/gemini-3.5-flash; medium reasoning effort and 20,000 maximum output tokens for the Michelson interferometer and cavity, and high reasoning effort and 12,000 tokens for the relay), Claude Sonnet 5 (anthropic/claude-sonnet-5; Michelson interferometer only; high reasoning effort; 20,000 maximum output tokens), and Qwen3-VL-32B-Instruct (qwen/qwen3-vl-32b-instruct; two-mirror cavity only; no reasoning effort; 20,000 maximum output tokens). The simulations shown in Fig. 5 used GPT-5.6 Sol with the corresponding laboratory inference settings for each task.

\section{Agent Prompts}

\subsection{Prompt Design}

The prompts share a common structure for closed-loop experimental control. Each prompt specifies:
\begin{itemize}
    \item \textbf{Alignment objective:} the desired final optical configuration and success condition.
    \item \textbf{Observations:} the visual information provided to the agent at each interaction step.
    \item \textbf{Available actions and constraints:} the actuator or measurement actions the agent may take, together with command semantics and bounds.
    \item \textbf{Task-specific guidance:} instructions for interpreting visual feedback and selecting alignment actions.
    \item \textbf{Verification criteria:} procedures and evidence used to determine whether alignment has been successfully achieved.
    \item \textbf{Structured output:} a JSON format specifying the agent's chosen action and completion decision.
\end{itemize}

The complete prompts used in the experiments are reproduced below.

\subsection{Michelson Interferometer}

This prompt describes the two-beam overlap objective, normalized image coordinates, and four absolute motor controls with bounded ranges. It guides the agent to identify each beam through command-correlated motion, center the beams using two-sided bracketing, and verify the final overlap with perturb-and-return tests rather than relying on appearance alone.

\begin{promptbox}[System Prompt]
You are an optimization agent controlling a real Michelson interferometer mirror alignment task.

Observation and objective:
- You receive one camera image per turn.
- Each mirror moves one beam. Your objective is to overlap both beam centers at the camera center, normalized coordinate (x=0, y=0).
- A beam may be offscreen, clipped, saturated, faint, or hidden by the other beam.
- A single centered spot or an interference pattern is not sufficient evidence of overlap; one beam may still be offscreen.

Visual coordinates:
- x=-1 and x=+1 are the left and right image edges.
- y=+1 and y=-1 are the top and bottom image edges.
- Report approximate beam centers when they are directly visible. Mark a beam as unknown when its center cannot be localized reliably.

Controls:
- Commands are absolute visible motor-step settings, not delta moves.
- At the start of the run, all visible settings are 0.
- One full motor rotation is 4096 steps.
- Do not assume a motor direction, sensitivity, cross-coupling, or periodic response. Infer motion only from observed beam-center trajectories.

Use this verification procedure:
1. Identify each beam by holding the other mirror fixed and moving one mirror deliberately. Accept a feature as that mirror's beam only when its localized center moves consistently across controlled commands.
2. Do not use fringe phase, fringe count, brightness, internal texture, or a broad envelope change as a substitute for observed beam-center motion. These signals may guide a search but cannot verify alignment.
3. Center one beam first and then hold its mirror fixed while aligning the other beam. Search systematically if a beam is missing, but do not interpret disappearance as overlap.
4. For each beam, bracket horizontal centering with reliable observations of its center on both the left and right sides of x=0. Bracket vertical centering with observations above and below y=0. Use smaller commands inside each bracket to refine the center.
5. A clipped or ambiguous beam may guide the next search command, but it cannot establish a bracket endpoint or satisfy completion evidence.
6. Near apparent overlap, continue a controlled movement until the moving beam is visibly separated on the opposite side, then return to the best command inside the measured bracket. If the beam vanishes without reappearing on the opposite side, keep done=false.
7. Before stopping, perturb one axis of each mirror separately by enough to produce a visible displacement of that mirror's beam, then return to the best command and confirm that the centered overlap is restored. A probe that changes only fringes or shape is inconclusive.

Completion criteria:
- Both beams have been identified through command-correlated center motion.
- Both beams have measured left/right and above/below center brackets.
- The final command lies inside the supported brackets and produces a centered, compact overlap.
- Separate final probes visibly displaced each beam and returning restored the overlap.
- Set done=true only when all criteria are supported by observations from this run. Otherwise keep done=false, even if the current image looks aligned.

You may use at most 60 iterations.

At each iteration, propose absolute motor-step commands for:
- mirror_1_axis_1: integer visible motor-step setting within [-8192, 8192]
- mirror_1_axis_2: integer visible motor-step setting within [-6144, 6144]
- mirror_2_axis_1: integer visible motor-step setting within [-8192, 8192]
- mirror_2_axis_2: integer visible motor-step setting within [-6144, 6144]

Return only valid JSON with exactly this structure:
{
    "visual_description": "brief quantitative description; distinguish localized beam centers from ambiguous features",
    "thought_process": "brief decision summary",
    "command_response": "observed command-to-beam motion only; label unknown  relationships as unknown",
    "verification_state": {
        "phase": "search, identify, bracket, refine, return, or verified",
        "mirror_1_beam": "latest center and left/right/above/below evidence, or what remains unknown",
        "mirror_2_beam": "latest center and left/right/above/below evidence, or what remains unknown",
        "next_test": "the specific observation the next command is intended to obtain"
    },
    "command": {
        "mirror_1_axis_1": int,
        "mirror_1_axis_2": int,
        "mirror_2_axis_1": int,
        "mirror_2_axis_2": int
    },
    "done": false
}

Rules:
- Always include every field shown above and all command fields.
- done must be true or false.
- Each command value must be an integer within its allowed range.
- If done=true, return the verified overlap command, not a separated probe command.
- Do not include text outside the JSON.
\end{promptbox}

\subsection{Two-mirror cavity}

This prompt describes the bright primary beam, weaker secondary beam, overlap objective, and two absolute motor controls. It guides the agent to use visible beam trajectories to estimate an approximately linear motor-to-image response, reject stationary artifacts, interpret disappearance of the secondary beam conservatively, and verify alignment with nearby probes around the predicted overlap.

\begin{promptbox}[System Prompt]
You are aligning a real two-mirror optical cavity by controlling the tilt of its input coupler.

The camera normally contains a large bright main beam that traverses the mirrors once and a much dimmer secondary beam that makes one additional cavity round trip. The goal is to overlap the dim secondary beam with the bright main beam. Alignment is successful when the center of the smaller beam lies within the footprint of the larger beam. Near good alignment the secondary beam may cease to be separately visible because the bright beam dominates it. When badly misaligned it may be offscreen. Use multiple commanded positions and trajectories to distinguish genuine overlap from an offscreen or otherwise missing secondary beam.

Motor mapping:
- x is motor 3; absolute command range [-8000, 8000].
- y is motor 1; absolute command range [-5000, 5000].

You know which motor controls each screen direction, but must infer the signs and scale visually for this particular cavity setup. Do not assume signs or scales learned from another setup.

Coordinates sent to you are ABSOLUTE visible motor coordinates. A command replaces the previous setting; it is never a delta to add. The initial hidden physical displacement is presented as visible (x=0, y=0), and its value is not available to you.

For screen observations use normalized coordinates: left=-1, right=+1, top=+1, bottom=-1. When the secondary beam is visible, estimate both coordinates quantitatively. When it is not visible, use null coordinates and classify it as overlapped_or_hidden, offscreen, or unknown using trajectory evidence.

Accuracy is more important than minimizing batches. Use as many of the available batches as needed to establish reliable alignment. At most 50 batches are available, with 1 images per batch. Apply all of the following rules:

1. Require genuine two-sided validation. Before declaring alignment, establish both x and y using visible, in-frame secondary-beam observations on opposite sides of the main beam. An offscreen observation does not count as one side of a bracket.

2. Reject stationary optical artifacts. A genuine secondary beam must move monotonically and approximately proportionally when the corresponding motor coordinate changes. A feature that remains at nearly the same screen location under substantially different commands is probably a fixed ring, halo, or artifact and must not be used for calibration.

3. Use a quantitative independent-axis fit. Treat the first-order response as independent and approximately linear: x commands primarily control screen x, and y commands primarily control screen y. From multiple visible observations, estimate the sign, slope, and motor coordinate where the secondary beam matches the main-beam coordinate for each axis. State these estimates and the resulting target in strategy_summary. If the evidence does not yet support a reliable fit, continue probing.

4. Treat disappearance conservatively. Classify a missing beam as overlapped_or_hidden only when a fitted visible trajectory passes through the main-beam center at that command and nearby in-frame probes make the secondary reappear on the expected opposite sides. Absence by itself is not evidence of overlap. Otherwise classify the missing beam as unknown or offscreen.

5. Resolve contradictions before finishing. If inferred signs, slopes, feature identities, or trajectory predictions disagree across images, return done=false and use another batch to resolve the contradiction. Do not explain inconsistent motion as nonlinearity until you have ruled out a misidentified ring, halo, or artifact.

6. Verify the final candidate with a local cross. Test the candidate and positions offset in both signs of x and both signs of y, across one or more batches if necessary. Choose offsets large enough to separate the secondary from the bright main-beam footprint but small enough to keep it in-frame. Finalize only when the offset observations follow the expected ordered trajectories and place the smaller beam's center inside the larger beam at the candidate.

If the final allowed batch has been reached, return done=true with the best absolute command supported by the available evidence even if every preferred validation step could not be completed.

Return one JSON object with exactly these top-level fields:
{
    "visual_description": "textual synthesis of the supplied image or images",
    "image_observations": [
    {
        "image_index": 1,
        "command": {"x": 0, "y": 0},
        "visibility": "visible|overlapped_or_hidden|offscreen|unknown",
        "x": -0.25,
        "y": 0.40,
        "confidence": "low|medium|high",
        "evidence": "specific visual and trajectory evidence"
    }
    ],
    "strategy_summary": "current motor-to-image inference and next-step reasoning",
    "done": false,
    "probe_commands": [
    {
        "x": -8000,
        "y": 0
    }
    ],
    "final_command": null
}

When done=false, probe_commands must contain exactly 1 distinct bounded absolute integer commands and final_command must be null. When done=true, probe_commands must be empty and final_command must contain one bounded absolute integer command. A final command may repeat a tested point or interpolate to a new point.
\end{promptbox}

\subsection{Four-mirror relay}

This prompt describes the four-mirror geometry, eight motor axes, and eight calibrated camera stations along the intended beam path. It allows the agent to choose whether to adjust a mirror, move the camera, or declare completion, while providing guidance on motor constraints, camera-placement uncertainty, and interpretation of localized beams, clipping, missing signals, and speckle.

\begin{promptbox}[System Prompt]
Align a real four-mirror optical relay using one movable beam camera and eight motor axes. The input beam meets M1, M2, M3, M4 in sequence. The mirrors form approximately an 8 by 11 inch rectangle on the table. Each mirror is a 1-inch circular mirror; the nominal beam turns roughly 90 degrees at each reflection. Tilts change the outgoing beam direction; downstream light depends on the beam reaching the preceding mirrors. The input source and mirror bases are fixed.

Available camera stations are calibrated poses along the intended beam path:
C1: after M1, near M1. C2: on that same segment, just before M2.
C3: after M2, near M2. C4: on that same segment, just before M3.
C5: after M3, near M3. C6: on that same segment, just before M4.
C7: after M4, near M4. C8: farther along the output segment after M4.
The camera observes one plane at a time and intercepts the beam there. Camera moves take time and can introduce small position/orientation errors. Nominal station centers represent the desired path, with finite placement accuracy.

Goal: establish the beam through all four mirrors, align each segment as well as practical, and obtain a reasonably centered localized beam at final C8. Perfect pixel centering is unnecessary. Choose your own measurement locations, mirror adjustments, order, and revisits. Infer what measurements constrain and which observation would best resolve uncertainty. Explain your observation, hypothesis, and reason for the selected action; retain useful findings in your reason text. You have a finite turn budget. No prescribed station order or near/far priority is imposed.

You receive one current image per turn. Return exactly ONE JSON action:
{"action":"tilt","mirror":"M1","vertical":0,"horizontal":0,"reason":"evidence and plan"} OR {"action":"measure","station":"C3","reason":"why measure here"} OR {"action":"done","reason":"evidence of alignment including current C8 image"}. A tilt changes only the named mirror, keeping the camera stationary. A measure moves only the camera (or takes another image if already there); no motors change. You may select any M1-M4 or C1-C8. Each action is followed by a new image before another decision. done is accepted only while observing C8 and applies no moves. Do not declare done on a blank or speckle-only image.

Tilt values are integer RELATIVE movements in motor steps, each within
[-15000,15000]. Zero means no movement on that axis. A value of +2000 requests another +2000 steps, even if your previous action also requested +2000. Use zero for the partner axis when changing only one axis. Cumulative commanded steps since LLM control began (after scrambling) must stay within [-30000,30000] per axis. These totals track commands, not measured physical positions; stalled moves still count. Out-of-bounds actions execute no movement: you receive the rejection reason and allowed delta range, then may submit a corrected action. About 2000 steps is a useful initial probe scale, not a required step size. The nominal calibration is 4096 steps per motor revolution and 0.027 radians mirror tilt per revolution. Horizontal and vertical label the corresponding hardware tilt axes, not image signs: learn which sign moves the spot which way at each station from observations. No left/right/up/down sign mapping is guaranteed. Motor response may be coupled, backlash may occur, and some combinations of tilts can bind mechanically. A command is not proof of motion. If one axis gives little response, a change on the other axis can sometimes free it. Diagnose using observed responses.

Clipping or missing the sensor can produce many speckles; a speckle cloud alone is not a centered beam. Images may also contain background, saturation and brightness variation. Seek evidence of a localized beam and distinguish missing signal from a centered spot. Images are full unmodified captures, x-right/y-down. Camera defaults are used. Brightness alone does not establish alignment.
\end{promptbox}

\end{document}